\documentstyle[preprint,aps]{revtex} 
\input epsf

\newcommand{\ptMstar}{\mbox{$\tilde{p}^\star_{\cal M}(x)$}}

\newcommand{\pLM}{\mbox{$p_L({\cal M})$}}

\begin{document} 

\title{Chain length dependence of the polymer-solvent\\ critical point parameters}
\author{N.B. Wilding, M. M\"{u}ller and K. Binder}
\address{Institut f\"{u}r Physik, Johannes Gutenberg Universit\"{a}t, \\
Staudinger Weg 7, D-55099 Mainz, Germany.}

\date{October 1995}
\setcounter{page}{0}
\maketitle 

\begin{abstract}

We report grand canonical Monte Carlo simulations of the critical point
properties of homopolymers within the Bond Fluctuation model.  By employing
Configurational Bias Monte Carlo methods, chain lengths of up to $N=60$
monomers could be studied.  For each chain length investigated, the critical
point parameters were determined by matching the ordering operator
distribution function to its universal fixed-point Ising form.  Histogram
reweighting methods were employed to increase the efficiency of this
procedure.  The results indicate that the scaling of the critical
temperature with chain length is relatively well described by Flory theory,
i.e.  $\Theta-T_c\sim N^{-0.5}$.  The critical volume fraction, on the other
hand, was found to scale like $\phi_c\sim N^{-0.37}$, in clear disagreement
with the Flory theory prediction $\phi_c\sim N^{-0.5}$, but in good
agreement with experiment.  Measurements of the chain length dependence of
the end-to-end distance indicate that the chains are not collapsed at the
critical point. 

\end{abstract}
\thispagestyle{empty}
\begin{center}
PACS numbers 61.25.Hq, 64.70.Fx, 05.70.Jk
\end{center}
\newpage
\tighten
\section{Introduction and overview} 
\label{sec:intro}

\begin{figure}

\vspace*{0.5 in}
\setlength{\epsfxsize}{13cm}
\centerline{\mbox{\epsffile{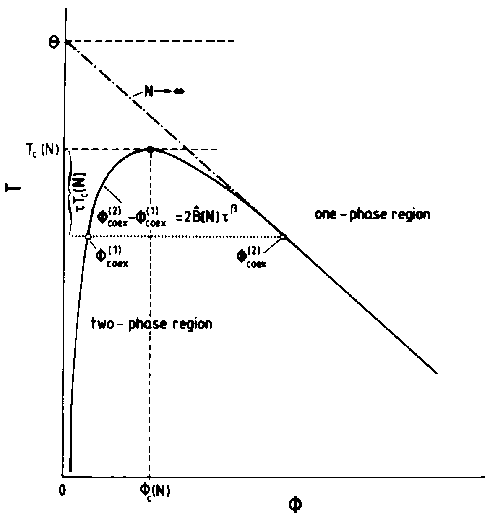}}} 

\caption{Schematic phase diagram of a polymer solution in the space of
the temperature $T$ and the volume fraction $\phi$. The ceoxistence
curve separates a dilute solution of collapsed chains [at
$\phi^{(1)}_{coex}$] from a semildilute solution of overlapping chains
[at $\phi^{(2)}_{coex}$]. These two branches of the coexistence curve
merge at a critical point $T_c(N), \phi_c(N)$. For $N\rightarrow\infty$
the critical point merges with the $\Theta$-point of a dilute polymer
solution [$T_c(N\rightarrow\infty)\rightarrow\Theta,\hspace{2mm}
\phi_c(N\rightarrow\infty)\rightarrow 0$] and the unmixing transition
has a tricritical character. At $T=\Theta$, the chain configurations
are ideal Gaussian coils, while their structure at $T_c(N)$ is
non-trivial.}

\label{fig:schem}
\end{figure}

When long flexible polymers are dissolved in a bad solvent there
exists a critical temperature $T_c(N)$ of unmixing slightly beneath
the $\Theta$-temperature (figure~\ref{fig:schem}). At this critical
temperature, the system phase separates into a very dilute (solvent
rich) solution of collapsed chains and a semidilute (polymer rich)
solution. The process is qualitatively described by the mean field
theory of Flory \cite{FLORY}, which predicts simple power laws for the
chain length ($N$) dependences of $T_c(N)$ and the corresponding
critical volume fraction $\phi_c(N)$:

\begin{eqnarray}
T_c(N) & = &\Theta/(1+1/\sqrt{N})^2 \\ \nonumber
&\approx &\Theta -2\Theta/\sqrt{N}, \hspace{3mm} N\rightarrow\infty
\label{eq:mftdep} 
\end{eqnarray}

\begin{eqnarray}
\phi_c(N)&= & 1/(1+\sqrt{N})\\ \nonumber
&\approx & 1/\sqrt{N}, \hspace{3mm} N\rightarrow \infty.
\label{eq:mfvfdep}
\end{eqnarray}
Another power law is predicted for the shape of the coexistence curve
near $T_c(N)$:

\begin{equation}
\phi^{(2)}_{coex}-\phi^{(1)}_{coex}=2\hat{B}(N)\tau^\beta,\hspace{2mm}\tau\equiv 1-T/T_c(N),
\label{eq:coexsh}
\end{equation}
with a critical order parameter exponent $\beta$ and a chain length
dependent critical amplitude $\hat{B}(N)$ given by
\begin{equation}
\beta=\beta_{MF}=1/2, \hspace{2mm} \hat{B}(N)\propto N^{-1/4}.
\label{eq:exponents}
\end{equation}
Further power laws describe the intensity of critical scattering, the
associated correlation lengths and the interfacial tension etc
\cite{STANLEY,WIDOM}, but will not be considered here. 

Notwithstanding the qualitative correctness of the Flory theory in
predicting a phase separation, it should be emphasised that the exponent
$\beta=\beta_{MF}=1/2$ in equation~\ref{eq:exponents}, as well as the
powers of $N$ in equations~\ref{eq:mftdep}--\ref{eq:exponents} are mean
field results, and thus cannot be expected to be quantitatively correct.
More generally one expects that (we follow the notation of a recent
experimental study \cite{ENDERS})

\begin{eqnarray}
\hat{B}(N)&\propto &N^{-x_1},\label{eq:Ndep} \\ \nonumber
 \phi_c(N)&\propto &N^{-x_2}, \\ \nonumber
 \Theta-T_c(N) & \propto &N^{-x_3},
\end{eqnarray}
where the mean field values of the exponents defined in
equation~\ref{eq:Ndep} are

\begin{eqnarray}
x_1^{MF}&=& 1/4,\\ \nonumber
x_2^{MF}&=& x_3^{MF}=1/2.
\end{eqnarray}
It is an interesting open question to ask what are the correct values of
these exponents. While it is generally accepted from the
``universality principle''\cite{FISHER}, as well as experimental
findings \cite{ENDERS,PERZYNSKI,DOBASHI,SHINOZAKI,SANCHEZ1,CHU,SANCHEZ,XIA},
that the phase separation of polymer solutions falls in the same
universality class as the three dimensional Ising model, so that
\cite{GUILLOU}

\begin{equation}
\beta \approx 0.325,
\label{eq:isbeta}
\end{equation}
the theoretical understanding of the exponents $x_1,x_2,x_3$ in
equation~\ref{eq:Ndep} is rather limited. Experimental data
have yielded the estimates 
\cite{ENDERS,PERZYNSKI,DOBASHI,SHINOZAKI,SANCHEZ1,CHU,SANCHEZ,XIA,IZUMI}

\begin{eqnarray}
x_1 &\approx & 0.23-0.34,\label{eq:experiment}\\  \nonumber
x_2 &\approx & 0.38 \\ \nonumber
x_3 &\approx & 0.47-0.50.
\end{eqnarray}
However, theoretical estimates for these exponents are still
controversial. De Gennes \cite{DEGENNES} suggested that in the limit
of large $N$, one has the same scaling behaviour as in mean field
theory, i.e. the coexistence curve scales as

\begin{equation}
\phi_{coex}^{(2)}-\phi_{coex}^{(1)}= \frac{1}{\sqrt{N}}\tilde{\phi}(\sqrt{N}\tau),
\label{eq:degennes}
\end{equation}
Since $\phi(z)$ must behave for small argument $z$ as
$\tilde{\phi}(z)\propto z^\beta$, this yields 
\begin{equation}
x_1=(1-\beta)/2\approx 0.34,
\end{equation}
which is roughly compatible with experiments. However, the scaling with
$\sqrt{N}$ in equation~\ref{eq:degennes} implies that $x_2=1/2$, which
clearly disagrees with equation~\ref{eq:experiment}. Muthukumar \cite{MUTHU} on the other
hand, suggested that in a limit where ternary interactions are
important, one should have different exponents, namely

\begin{equation}
x_1=x_3=2/9, \hspace{2mm} x_2=1/3.
\label{eq:MUTHU}
\end{equation}

Subsequently this problem has received further attention in the
literature \cite{WIDOM,STEPANOW,KHOLODENKO,CHERAYIL,LHUILLIER}. Recall
that the scaling structure in equation~\ref{eq:degennes} can be
justified in terms of a Ginzburg criterion \cite{GINZBURG} if one
assumes that the chain linear dimensions are ideal \cite{DEGENNES}.
It then follows (remembering that the chain gyration radius enters as
a critical amplitude prefactor in the mean-field power laws of the
correlation length of the monomer density fluctuations), that the
critical density $\phi_c(N)$ coincides (up to a universal prefactor)
with the onset of the ``semi-dilute'' regime, where chains overlap
significantly \cite{DEGENNES}. This assumption is plausible because of
the vicinity to the $\Theta$-state ($T=\Theta, \phi\rightarrow 0$),
where chains indeed behave ideally and the gyration radius scales as

\begin{equation}
R_g \propto N^{1/2},\hspace{2mm} T=\Theta, \hspace{2mm}\phi\rightarrow 0, \hspace{2mm}N\rightarrow\infty.
\label{eq:Rg1}
\end{equation}
However the fact that for $T<\Theta$ and $\phi\rightarrow 0,
N\rightarrow \infty$ chains are collapsed:

\begin{equation}
R_g \propto N^{1/3}, \hspace{2mm} T < \Theta, \hspace{2mm}\phi\rightarrow 0, \hspace{2mm} N\rightarrow\infty,
\label{eq:Rg2}
\end{equation}
implies that one does not really know how $R_g$ scales with $N$ at the
critical point. Therefore it is tempting to generalize the scaling
ansatz~\ref{eq:degennes} as follows \cite{CHERAYIL}.

\begin{equation}
\phi_{coex}^{(2)}-\phi_{coex}^{(1)}= \frac{1}{N^{x_4}}\tilde{\phi}(N^{x_4}\tau).
\label{eq:cherayil}
\end{equation}

Equation~\ref{eq:experiment} is, of course,  still consistent with the behaviour of the
coexistence curve at fixed $\tau$ in the limit $N\rightarrow\infty$
\cite{WIDOM}:

\begin{equation}
\phi_{coex}^{(1)}=0, \hspace{2mm}\phi_{coex}^{(2)}=\frac{3}{2}(1-T/\Theta),
\end{equation}
if $\tilde{\phi}(z\rightarrow\infty)=\frac{3}{2}z$. Since
for small $z=N^{x_4}\tau$, the scaling function must behave as 
$\tilde{\phi}(z)\propto z^\beta$ in order to comply with
equations~\ref{eq:coexsh} and \ref{eq:isbeta}, we conclude that 
\begin{equation}
\phi_{coex}^{(2)}-\phi_{coex}^{(1)}=N^{x_4(1-\beta)}\tau^\beta, \hspace{2mm} x_1=x_4(1-\beta).
\label{eq:cher1}
\end{equation}
From renormalization group arguments, Cherayil \cite{CHERAYIL}
has suggested that the exponents $x_2$ and $x_3$ can be expressed in terms
of the new exponent $x_4$ as 

\begin{equation}
x_2=1-x_4, \hspace{2mm}x_3=x_4.
\label{eq:cher2}
\end{equation}
This theory, however, does not yield a prediction for $x_4$ itself,
and to fit some experimental data it was assumed that $x_4=0.62,
x_2=0.38$ \cite{CHERAYIL}. Kholodenko and Qian \cite{KHOLODENKO} have
presented arguments that the exponent $x_2$ is not even a universal
quantity. If the scaling relations of Cherayil
(equation~\ref{eq:cher1},\ref{eq:cher2}) hold, this would imply that
$x_1, x_2$ and $x_3$ are all system specific quantities, depending
upon the material under consideration! Finally, we note that
Muthukumar's result, equation~\ref{eq:MUTHU}, disagrees with the above
scaling relation $x_1=x_3(1-\beta)$, and thus the theoretical
situation is clearly somewhat confusing.

In view of these problems and the difficulties of extracting all
relevant information from experiments (one not only wishes to check
the relations of equation~\ref{eq:Ndep} but also seeks to clarify how
the chain span scales with $N$ at criticality), study of this problem
by Monte-Carlo computer simulations techniques \cite{BINBOOK} is
highly desirable. In fact there has been some previous work on this
problem which considered the vapour-liquid phase diagram of alkane
chains \cite{SMIT} and coarse-grained off-lattice polymer models (see
e.g.  \cite{SHENG,ESCOBEDO}). However the work of reference \cite{SMIT}
considers the problem of estimating absolute values of $T_c(N)$ and
$\phi_c(N)$ for a chemically realistic model of alkanes for small $N$
and does not address the universal properties of the limit
$N\rightarrow\infty$. The Gibbs ensemble Monte-Carlo method of
Panagiotopoulos \cite{PANAGIO,MACKIE} allows an efficient
estimation of the coexistence curve well below the critical point, but
a precise estimation of critical point parameters is difficult in this
framework.

An alternative approach for estimating critical point properties from
simulations is based on finite-size scaling \cite{PRIVMAN,BINDER1}. This
approach has been very successful for both symmetrical \cite{DEUTSCH1}
and asymmetrical \cite{DEUTSCH2} polymer mixtures in conjunction with the
bond-fluctuation lattice model \cite{BFM} and semi-grand canonical
ensemble simulation techniques \cite{SARIBAN}. These studies also relied
on the use of histogram reweighting \cite{FERRENBERG} and (in the
asymmetric case) recent advances in disentangling order parameter and
energy fluctuations near criticality in a finite-size scaling context
\cite{WILDING}. 

In the present work we attempt to apply a related approach to study the
liquid-vapour critical point of homopolymers within the Bond Fluctuation
model. This problem is, however, somewhat more intricate than that of
polymer mixtures since one must employ the grand canonical ensemble
(GCE) \cite{WILDING} in order to effectively deal with the strong
near-critical density fluctuations.  As is well known, GCE simulations
for chain molecules are extremely difficult, since the insertion
probability for a polymer chain into a many chain systems is vanishingly
small \cite{BINBOOK,FRENKEL,SIEPMANN,FRENKEL1}. For chains that are not
too long (and/or systems that are not too dense), this problem can be
eased by the Configurational Bias Monte-Carlo (CBMC) Method
\cite{FRENKEL,SIEPMANN,FRENKEL1}. In the present paper we combine CBMC
with histogram reweighting and a finite-size scaling analysis in its
form extended to asymmetric systems \cite{DEUTSCH2,WILDING}. By this
special combination of recent techniques (which will be briefly reviewed
in section~\ref{sec:back}) we are able, for the first time to obtain
accurate results, both for $\phi(N)$ and $T_c(N)$ up to $N=60$ effective
monomers. Since the effective bond in the bond-fluctuation model can be
thought of as corresponding to $3$ to $5$ chemical bonds (when a mapping
to chemically realistic chain molecules is attempted \cite{BFM}), our
simulations thus correspond to a degree of polymerization up to a few
hundred chemical bonds along the chain backbone.

Section~\ref{sec:results} then presents our results, including a
highly precise estimation of the $\Theta$ temperature from an analysis
of the gyration radius of simple isolated chains. We obtain both the
location of the critical point in the $(T,\phi)$ plane as a function
of chain length and, for the first time, the associated dependence of
the chain span. In section~\ref{sec:discuss} we discuss our results
and compare them to the theoretical ideas sketched above. We obtain
very good agreement with experiment, but as in the latter the need to
study much longer chains is clearly apparent to definitively clarify
the true asymptotic behaviour for chain lengths $N\rightarrow\infty$.

\section{Algorithmic and computational aspects}

\label{sec:back}

The bond-fluctuation model (BFM) studied in this paper is a
coarse-grained lattice-based polymer model that combines computational
tractability with the important qualitative features of real polymers,
namely monomer excluded volume, monomer connectivity and short range
interactions. Within the framework of the model, each monomer occupies
a whole unit cell of a 3D periodic simple cubic lattice.  Neighboring
monomers along the polymer chains are connected via one of $108$
possible bond vectors. These bond vectors provide for a total of $5$
different bond lengths and $87$ different bond angles. Thermal
interactions are catered for by a short range inter-monomer potential.
The cutoff range of this potential was set at $r_m=\sqrt{6}$ (in units
of the lattice spacing), a choice which ensures that the first peak of
the correlation function is encompassed by the range of the potential.
We note also, that within our model, solvent molecules are not
modelled explicitely, rather their role is play by vacant lattice
sites. Further details concerning the BFM can be found in reference
\cite{BFM}.

To implement a grand canonical ensemble simulation of the BFM, the
Configurational Bias Monte Carlo (CBMC) method was employed
\cite{FRENKEL,SIEPMANN,FRENKEL1}.  The CBMC scheme utilizes a biased insertion method
to `grow' a polymer into the system in a stepwise fashion, each
successive step being chosen so as to avoid excluded volume where
possible. For brevity we shall merely outline the GCE implementation of
this CBMC method and refer the reader to reference \cite{SMIT1} for a
fuller description.

Within the GCE scheme there are two complementary types of moves,
insertion attempts and deletion attempts, both of which are made with
equal frequency. An insertion move first involves attempting to grow a
candidate polymer into the system. The basic strategy for achieving
this is to insert successive monomers of the chain into the system one
by one. The position of each successive monomer is chosen
probabilistically from the set of $108$ possible BFM bond vectors
emanating from the previously inserted monomer.  The selection
probability for each of the possible monomer positions is weighted by
its Boltzmann factor, effectively biasing the choice in favour
of low energy chain configurations.  In order to keep track of the
accumulated bias, a book keeping scheme is maintained. Once a
candidate chain has been successfully grown, it is submitted to a
Monte-Carlo lottery to decide whether or not it is to be accepted. The
total chain construction bias is compensated for in the acceptance
probability, thereby ensuring that detailed balance is obeyed.

For chain deletion moves, one chooses a chain at random from those in
the system and `reconstructs' its bias by examining the alternative
growth scenarios at each step of the chain. The candidate chain for
deletion is also submitted to a Monte-Carlo lottery to decide whether
the proposed deletion should take place. As with the insertion
lottery, the chain bias is taken into account in the deletion
probability. The chemical potential, $\mu$, which controls the system
chain density, also enters into the acceptance probability for both
insertion and deletion.

The principal observables measured in the course of the simulations
were the monomeric volume fraction:

\begin{equation}
\phi=8nN/V
\end{equation}
and the dimensionless energy density:

\begin{equation}
u=8w^{-1}\Phi(\{r\})/V
\end{equation}
where $n$ is the number of chains, $\Phi(\{r\})$ is the
configurational energy, $w$ is the depth of the square well
interaction potential (so that $T=w^{-1}$) and $V$ is the system
volume.  Here the factor of $8$ derives from the number of lattice
sites occupied by one monomer in the BFM. Measurements of $\phi$ and
$u$ were performed at intervals of $500$ chain insertion attempts and
accumulated in the joint histogram $p_L(\phi,u)$. The final histograms
comprised some $10^5$ entries. Also measured were the distributions of the chain
radius of gyration and the chain end-to-end distances.

\begin{figure}

\vspace*{0.5 in}
\setlength{\epsfxsize}{19cm}
\centerline{\mbox{\epsffile{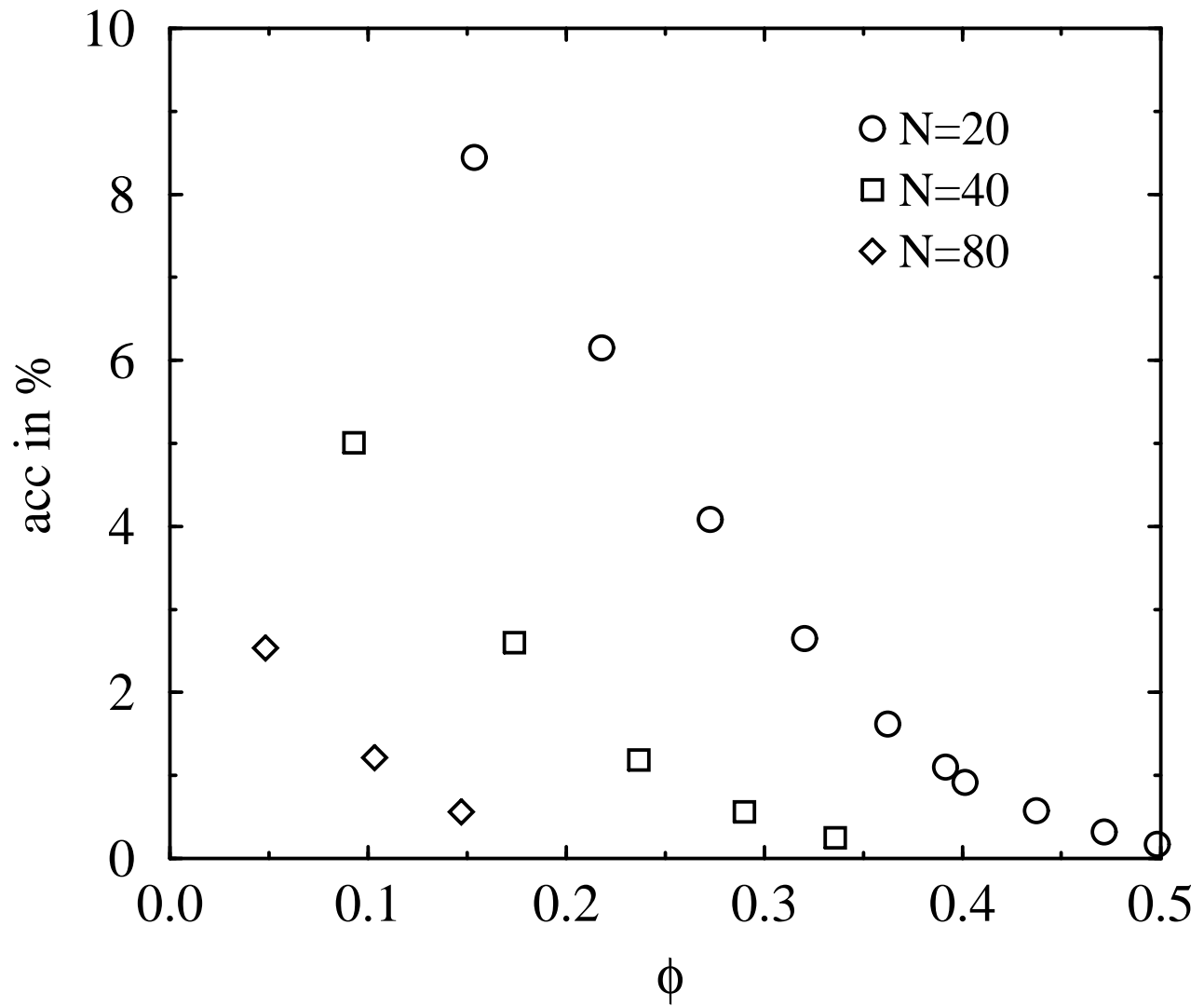}}} 
\caption{The CBMC acceptance rate as a function of the monomeric
volume fraction for chain lengths $N=20,40,80$.}

\label{fig:acc}
\end{figure}

Using the GCE algorithm, chains of lengths $N=10,20,40,60$ were
studied.  For the $N=10$ and $N=20$ system size $V=40^3$ and $V=50^3$
were employed, while for $N=40$ and $N=60$ chain lengths only the
$V=50^3$ was studied. Unfortunately it was not possible to study
chains longer than $N=60$ since the acceptance rate for chain
insertions falls exponentially with increasing $N$ and volume fraction
$\phi$. This problem is illustrated in figure~\ref{fig:acc}, where we
plot the acceptance rate for a number of chain lengths as a function
of the monomeric volume fraction. One sees for example, that for
$N=80$ the acceptance rate is too low to provide reliable statistics
within reasonable run times. Indeed, even for our longest chain length
$N=60$, extremely long runs were required to gather adequate
statistics.

Having outlined our model and simulation technique we now turn to a brief
description of our data analysis methods.  As mentioned in the introduction,
finite-size scaling (FSS) methods are an indispensable tool for the proper
treatment of critical behaviour, serving as they do to provide estimates
infinite-volume critical properties from simulations of finite-size systems. 
The FSS methods we shall employ here are especially tailored to fluid
systems and have been described in detail elsewhere \cite{WILDING}.  The
basic idea is to exploit the Ising character of the polymer liquid-vapor
critical point to accurately locate the critical point.  This is done by
observing that precisely at criticality the distributions of certain readily
measurable observables assume scale-invariant universal forms.  The
particular universal scaling form on which we shall focus, is the
distribution of the ordering scaling operator \pLM\ .  For the Ising model,
the special symmetry between the coexisting phases implies ${\cal
M}\rightarrow m$ (the magnetisation).  The critical point form of $p_L(m)$
is independently known from extensive studies of large Ising lattices
\cite{HILFER}.  For fluids, however, the lack of symmetry between the
coexisting phases implies \cite{WILDING} that the ordering operator is a
linear combination of the fluid density and energy density i.e.  ${\cal
M}\rightarrow \phi+su$, where $s$ is a system specific `field mixing'
parameter that controls the strength of the coupling between the density and
energy fluctuations. 

Thus, in principle, one is able to accurately locate the critical
point of a fluid system simply by tuning the $T,\mu$ and $s$ until the
measured form of $p_L({\cal M})$ matches the known fixed point Ising
form. In performing this task, the histogram reweighting method
\cite{FERRENBERG,DEUTSCH1} can be of great assistance. This technique
allows one to generate estimated histograms $p_L(\phi,u)$ for values
of the control parameters $T$ and $\mu$ other than those
at which the simulations were actually performed.  Such extrapolations
are generally very reliable in the neighborhood of the critical point,
due to the large critical fluctuations \cite{FERRENBERG}. In what
follows we shall detail the application of all these techniques to the
problem of determining the liquid-vapor critical point parameters of
our polymer model.

\section{Procedure and results}
\label{sec:results}

\begin{figure}

\vspace*{0.5 in}
\setlength{\epsfxsize}{19cm}
\centerline{\mbox{\epsffile{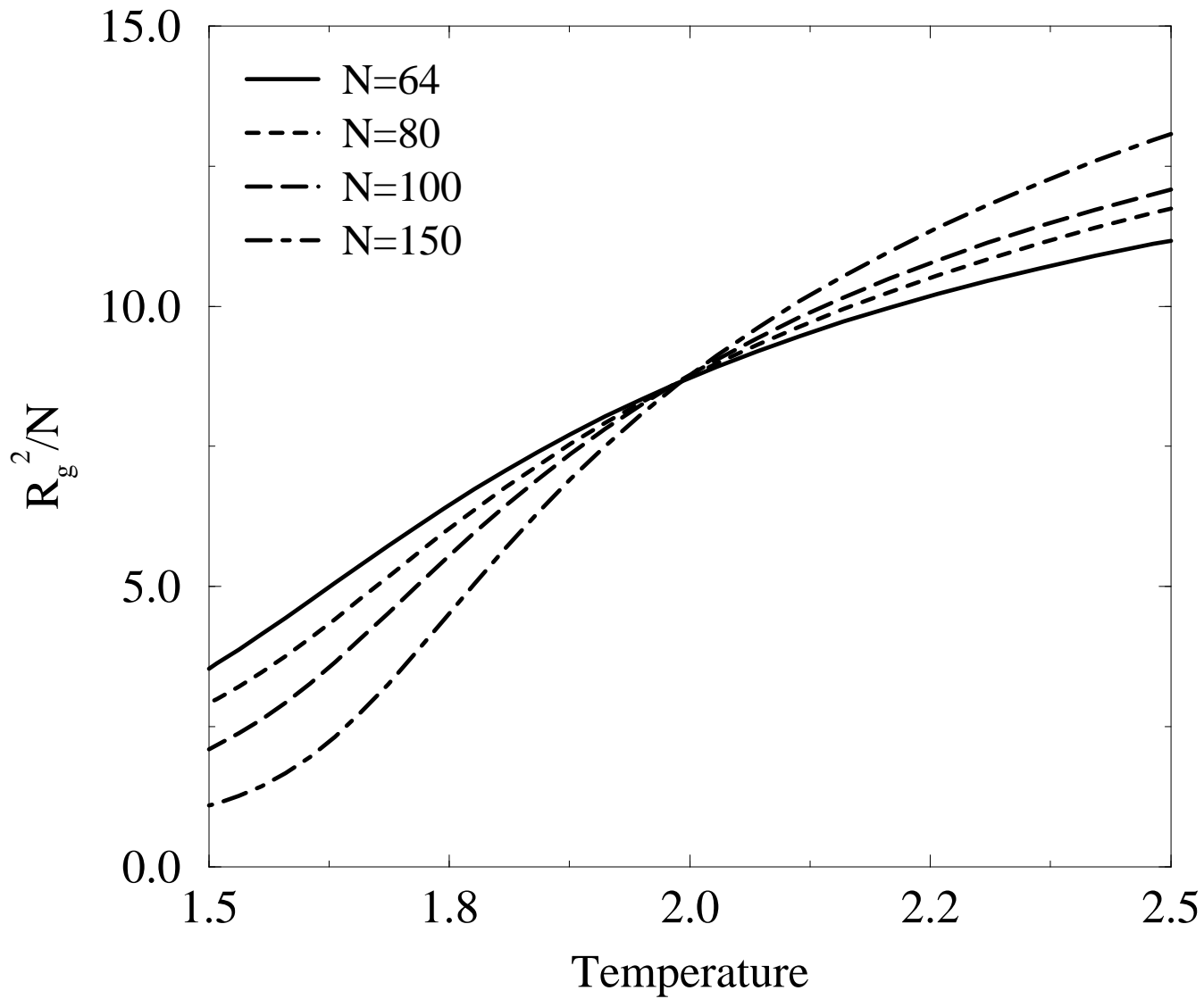}}} 

\caption{The temperature dependence of $R_g^2/N$ for a single chain
with a variety of chain lengths.  All simulations were performed at
the temperature $T=2.0$ and the temperature dependence obtained by
histogram extrapolation. The common intersection point for large $N$
yields the estimate $\Theta=2.020(3)$.}

\label{fig:theta}
\end{figure}

The first task undertaken was a determination of the
$\Theta$-temperature for our model, knowledge of which is a
prerequisite for studying the scaling of $\Theta-T_c(N)$ (cf.
equation~\ref{eq:Ndep}). To achieve this the gyration radius $R_g$ of
single chains was studied as a function of temperature and chain
length.  From equations~\ref{eq:Rg1} one sees that precisely at the
$\Theta$-temperature, and modulo corrections to scaling, $R_g^2/N$
should be independent of $N$. Extensive simulations were therefore
carried out for single chains of length $N=64,80,100$ and $150$ at a
temperature $T=2.0$. The full temperature dependence of $R_g$ for each
chain length was subsequently obtained by histogram reweighting.  This
involves recording the joint histogram of the gyration radius and
conformational energy of each configuration generated. The histogram
for other temperatures may then be obtained by reweighting the
Boltzmann factor for each histogram entry in the manner described in
reference \cite{FERRENBERG}. Figure~\ref{fig:theta} show the resulting
curves of $R_g^2/N(T)$, which exhibit a very precise intersection
point at $T=2.020(3)$, a value that we therefore adopt as our estimate
of the $\Theta$-temperature.
 
In general for fluid systems, the coexistence curve is not known {\em
a-priori} and must therefore be identified empirically as a prelude to
locating the critical point itself. In the following we exemplify the
general strategy for determining the critical parameters by
considering the case of the $N=20$ system.

\begin{figure}[h]
\vspace*{0.5 in}
\setlength{\epsfxsize}{10cm}
\centerline{\mbox{\epsffile{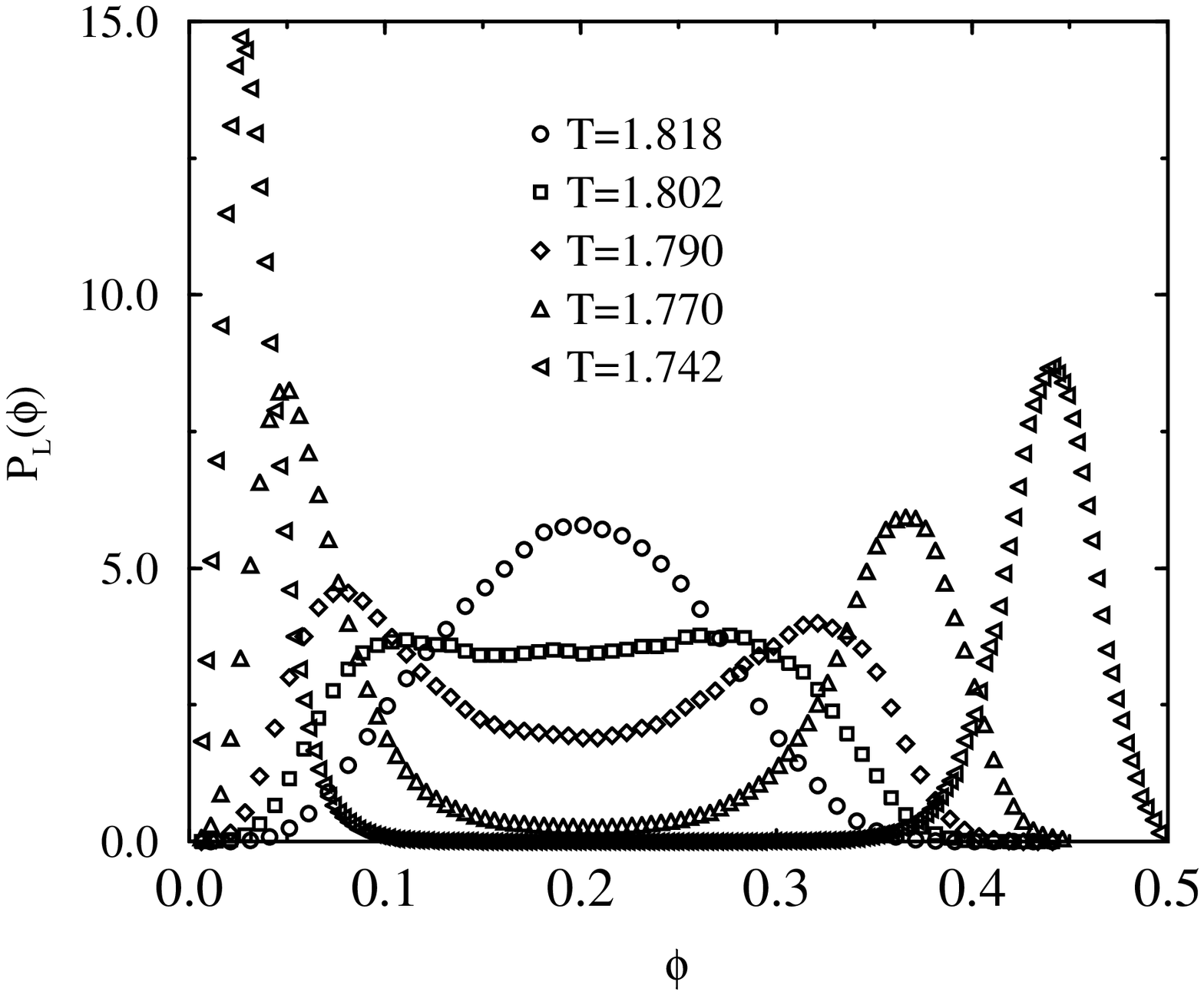}}} 
\centerline{\mbox{\epsffile{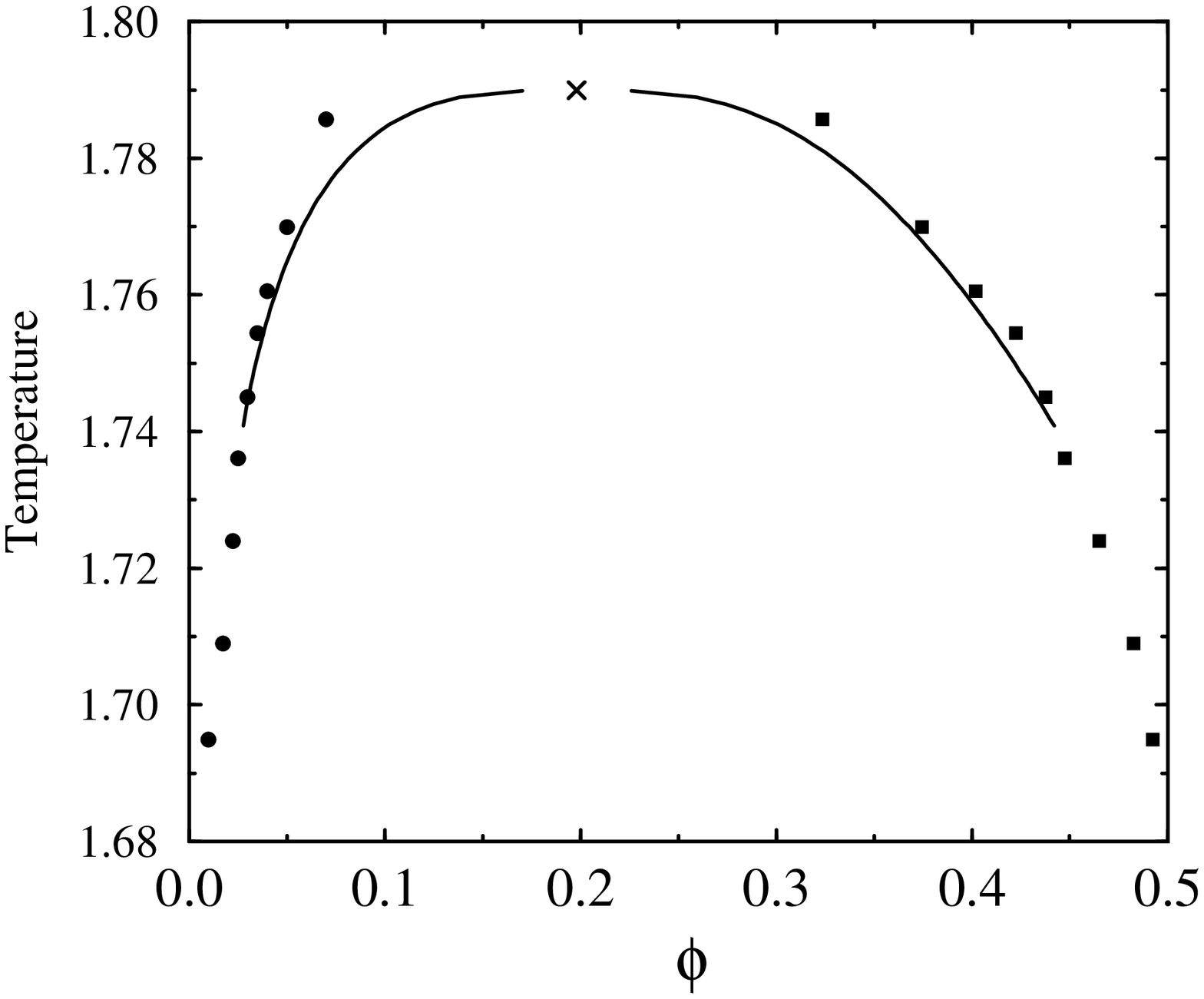}}} 

\caption{{\bf (a)} The distribution function of the monomeric volume
fraction $p_L(\phi)$ for the N=20 system at a selection of
temperatures along the line of liquid-vapor coexistence. {\bf (b)} The
measured liquid (squares) and vapour (circles) peak densities
corresponding to the histogram extrapolation along the coexistence
curve. Also shown is an attempted fit to the data (full line) and the
estimate for the critical point (X), both obtained as described in the
text.}

\label{fig:cxcurve}
\end{figure}

Initially a temperature of $T=1.75$ somewhat beneath the $\Theta$
temperature was chosen and the approximate value of the coexistence
chemical potential was determined by tuning $\mu$ until $p_L(\phi)$
exhibited a double peaked structure. A long run was then carried out
at this near-coexistence $\mu$ value, in which the histogram of
$p_L(\phi,u)$ was accumulated. A histogram extrapolation based on this
data was then used to extrapolate along the coexistence curve using
the equal peak-weight criterion for $p_L(\phi)$ \cite{EWR}. In this
way a sizeable portion of the the near-critical coexistence curve (and
its analytic extension\cite{WILDING}) could be located. Representative
forms of the density distributions along this line of
pseudo-coexistence are shown in figure~\ref{fig:cxcurve}(a), while the
positions of the density peaks are shown in figure~\ref{fig:cxcurve}(b).

\begin{figure}[h]
\vspace*{0.5 in}
\setlength{\epsfxsize}{19cm}
\centerline{\mbox{\epsffile{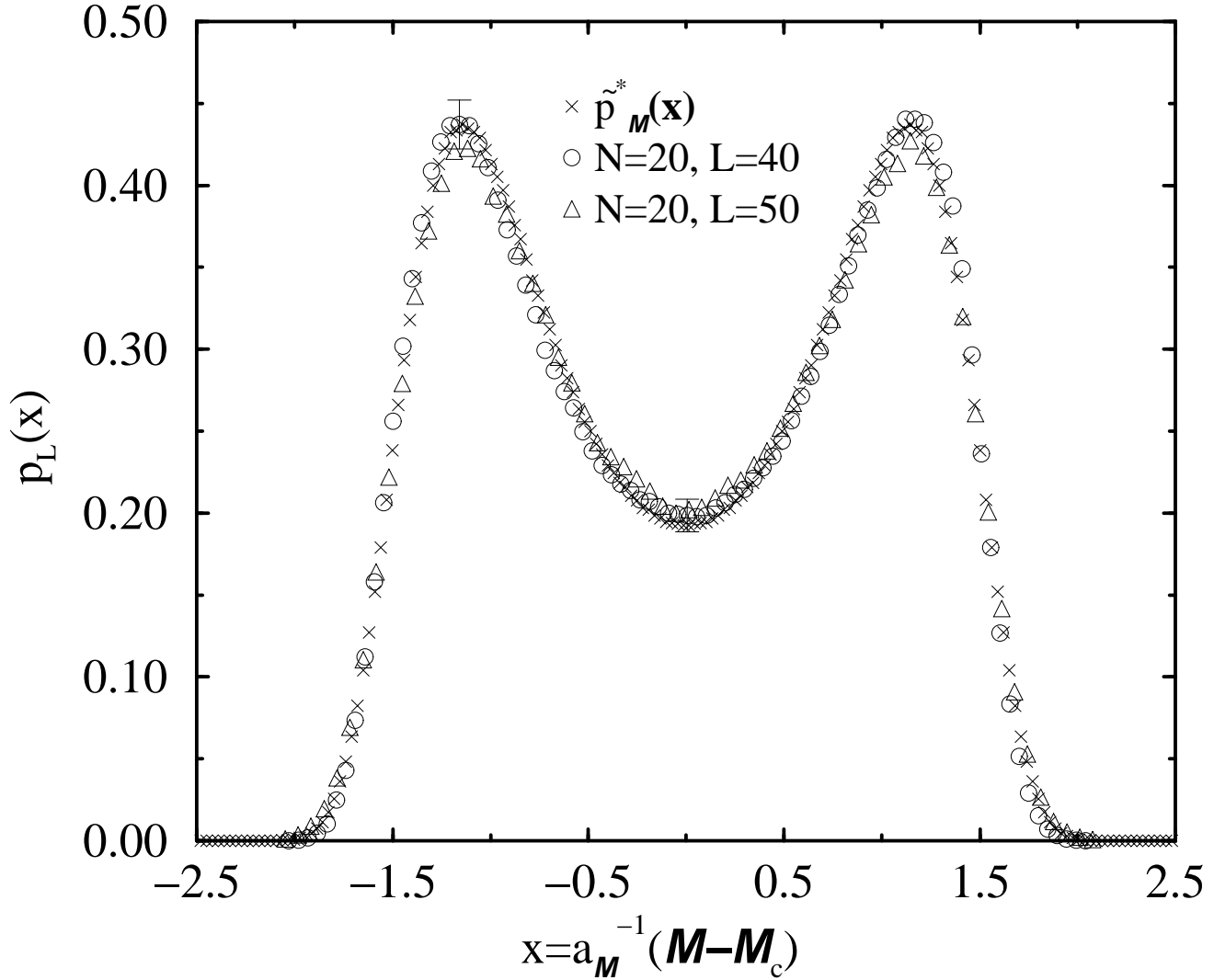}}} 

\caption{The ordering operator distribution \pLM\ of the polymer model
at the assigned critical parameters. Also shown for comparison is the
universal fixed point form \ptMstar\ obtained in a seperate study
\protect\cite{HILFER}. In accordance with convention all data has been scaled
to unit norm and variance, by choice of the scale factor $ a_{\cal
M}$.}

\label{fig:polyops}
\end{figure}

To locate the critical point along the line of phase coexistence, we
utilized the universal matching condition for the operator
distributions \pLM\ . Again applying the histogram reweighting
technique, the temperature, chemical potential and field mixing
parameter $s$ were tuned until the form of \pLM\ most accurately
matched the universal fixed point Ising form \ptMstar . The results of
performing this procedure are shown in figure~\ref{fig:polyops} for
the $V=40^3$ and $V=50^3$ system sizes. Given that these systems
contain on average less than $100$ polymer chains at criticality, the
quality of the data collapse is remarkable. The mappings shown were
effected for a choice of the parameters $T_c=1.791(5),
\mu_c=-5.16425(2), s=-0.135(4)$ where we have defined $\mu$ to be the
chemical potential per monomer. The associated estimates for the
critical volume fraction is $\phi_c=0.198(5)$. We note that this
estimate of $T_c$ is rather less than what would be obtained, were one
simply to extrapolate the liquid and vapour densities to the point at
which they merge [figure~\ref{fig:cxcurve}(b)].  Thus our results
further emphasise the finite-size errors that can arise using such a
procedure. 

We have also attempted to fit the coexistence curve data of
figure~\ref{fig:cxcurve}(b), using fits of the Ising form: $\phi_l =
\phi_c+0.745(T_c-T)+0.55(T_c-T)^{0.324}, \phi_v =
\phi_c+0.745(T_c-T)-0.55(T_c-T)^{0.324}$, with $T_c$ and $\phi_c$ assigned
the values obtained above.  For the Lennard-Jones fluid \cite{WILDING2}, a
fit of this type gave a good description of the coexistence data for
temperatures down to $0.8T_c$.  Here, one sees that a reasonable fit can be
obtained, but only for temperature within some $3\%$ of $T_c$.  This would
seem to suggest that the asymptotic Ising region in the polymer system is
much smaller than in the Lennard-Jones system. 

With regard to our estimates for the critical parameters, it should be
emphasized that they are subject to errors arising both from
corrections to scaling, as well as field mixing effects in the case of
the critical volume fraction \cite{WILDING}. While in principal one
can also correct for these effects, if one has access to a sufficient
range of system sizes and ample statistics (see eg. \cite{WILDING2}),
the computational difficulties of the present problem preclude such an
analysis.  Indeed, for the $N=60$ system studied here, only one system
size, $V=50$ was employed, this being the largest that could
reasonably be tackled.  Smaller system sizes were not studied since
these would contain so few chains at criticality as to be excessively
influenced by corrections to scaling. The matching to the universal
form for the largest size available was then the only guide to the
location of the critical point. Notwithstanding these problems
however, we feel on the basis of our experience of corrections to
scaling in other systems \cite{WILDING2}, that the quoted
uncertainties generously encompass the infinite-volume critical
parameters.

The procedure described above was repeated for the other chain lengths
studied, allowing estimates to be obtained for $T_c(N)$ and
$\phi_c(N)$.  However the computational difficulty became
progressively greater as $N$ increased, making the accumulation of
good statistics problematic. As a result it was not possible to
perform a reliable histogram extrapolation away from the critical
point into the sub-critical two-phase region and thus no information
on the $N$ dependence of the critical amplitude prefactor featuring in
equation~\ref{eq:Ndep} could be obtained. An additional hindrance to
probing the subcritical coexistence region is that the BFM appears to
be unable to support a liquid phase for volume fractions $\phi \agt
0.6$, instead collapsing into an amorphous crystal. This artifact is
traceable to the limited conformational entropy of our lattice-based
chains, and has also been observed in a previous study of tethered
chains using the same model \cite{LAI}.

\begin{figure}[h]
\vspace*{0.5 in}
\setlength{\epsfxsize}{19cm}
\centerline{\mbox{\epsffile{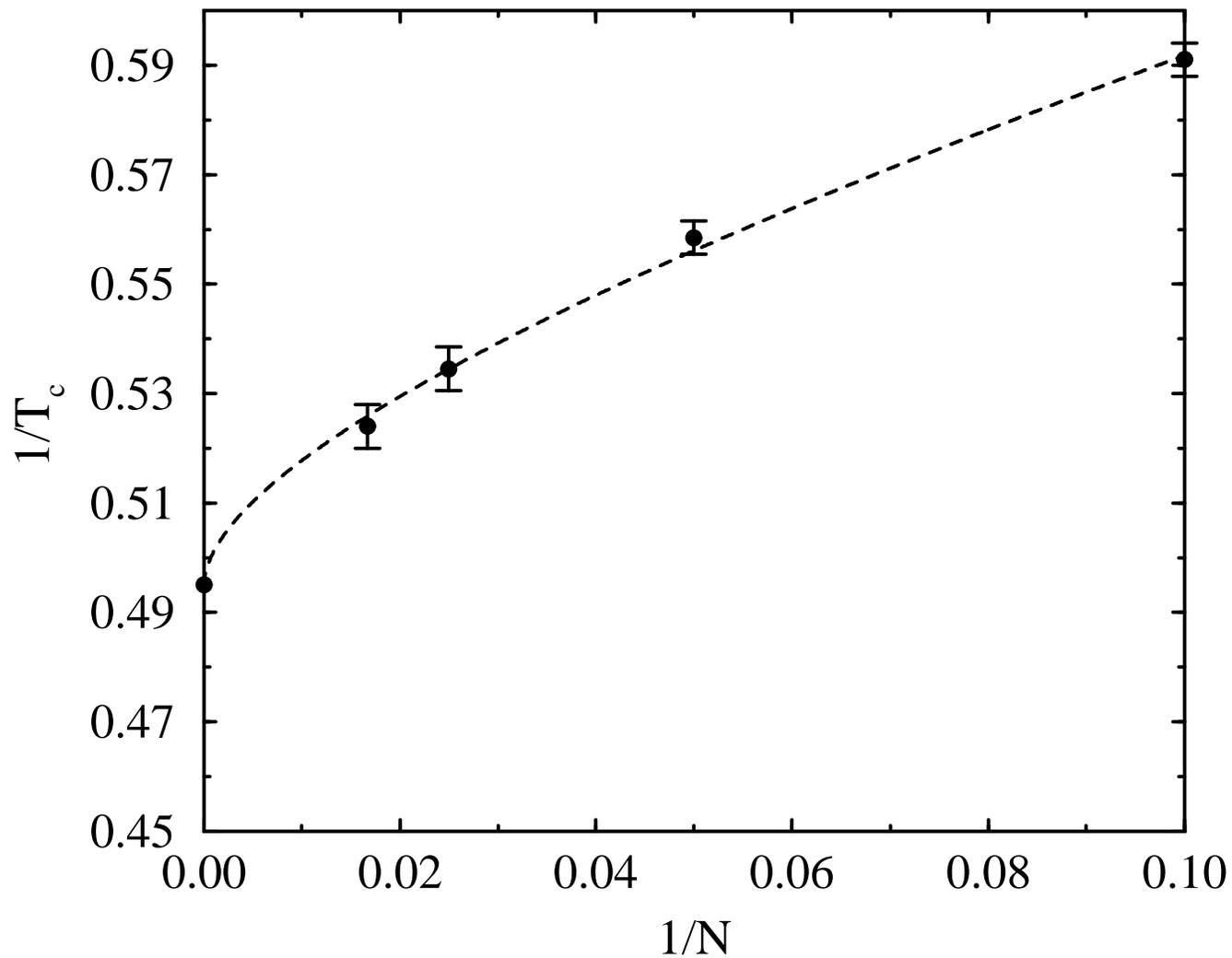}}} 

\caption{The measured estimates of the inverse critical temperature as
a function of inverse chain length, together with the $\Theta$
temperature, representing the critical temperature in the limit of
infinite $N$. Also shown is a fit of the form $1/T_c=1/\theta
(1+0.402N^{-0.5}+0.696N^{-1})$.}

\label{fig:TvsN}
\end{figure}

\begin{figure}[h]
\vspace*{0.5 in}
\setlength{\epsfxsize}{19cm}
\centerline{\mbox{\epsffile{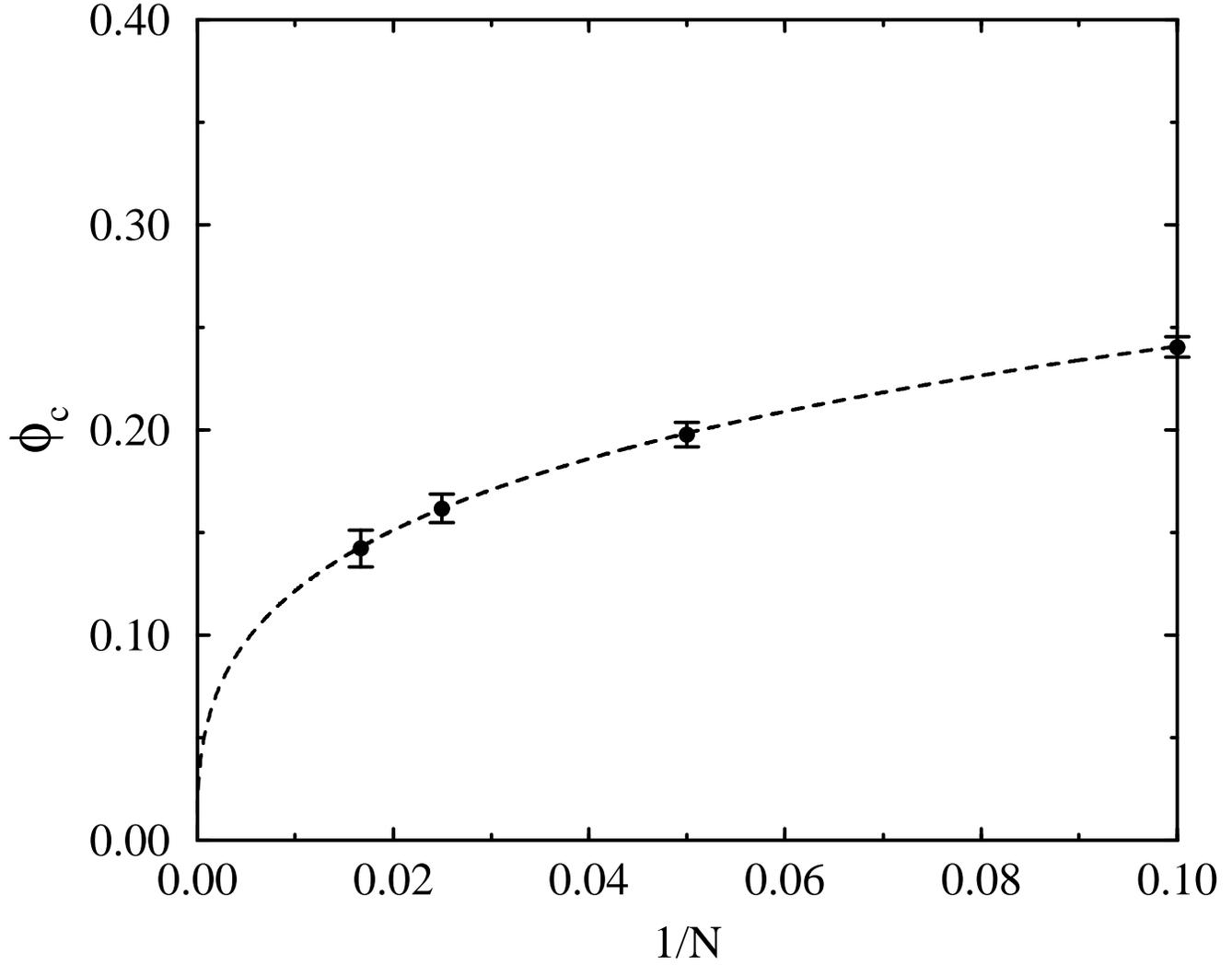}}} 

\caption{The measured estimates of the critical density as a
function of chain length, together with the infinite $N$ value $\phi_c=0$
Also shown is a fit of the form $\phi_c=(1.1126+1.3N^{0.369})^{-1}$}

\label{fig:DvsN}
\end{figure}

The results for $T_c(N)$ and $\phi_c(N)$ are plotted against
$N^{-1}$ in figures~\ref{fig:TvsN} and ~\ref{fig:DvsN}
respectively. For $T_c(N)$ we find that the data can be well fitted by
a Flory-type formula of the form $T_c=\Theta+a_1N^{-0.5}+a_2N^{-1}$
(where the $N^{-1}$ term can be thought of as a free-end correction and
where we {\em assume} that $lim_{N\to\infty}T_c=\theta)$,
although fits of the form $T_c(N)=\Theta+a_1N^{-x_3}$ yield a
comparable fit quality for values of $x_3$ in the range
$x_3=0.46$--$0.53$. For $\phi(N)$ we have performed a fit of the form
$\phi_c(N)=(b_1+b_2N^{x_2})^{-1}$, and obtain $x_2=0.37(2)$.

\begin{figure}[h]
\vspace*{0.5 in}
\setlength{\epsfxsize}{19cm}
\centerline{\mbox{\epsffile{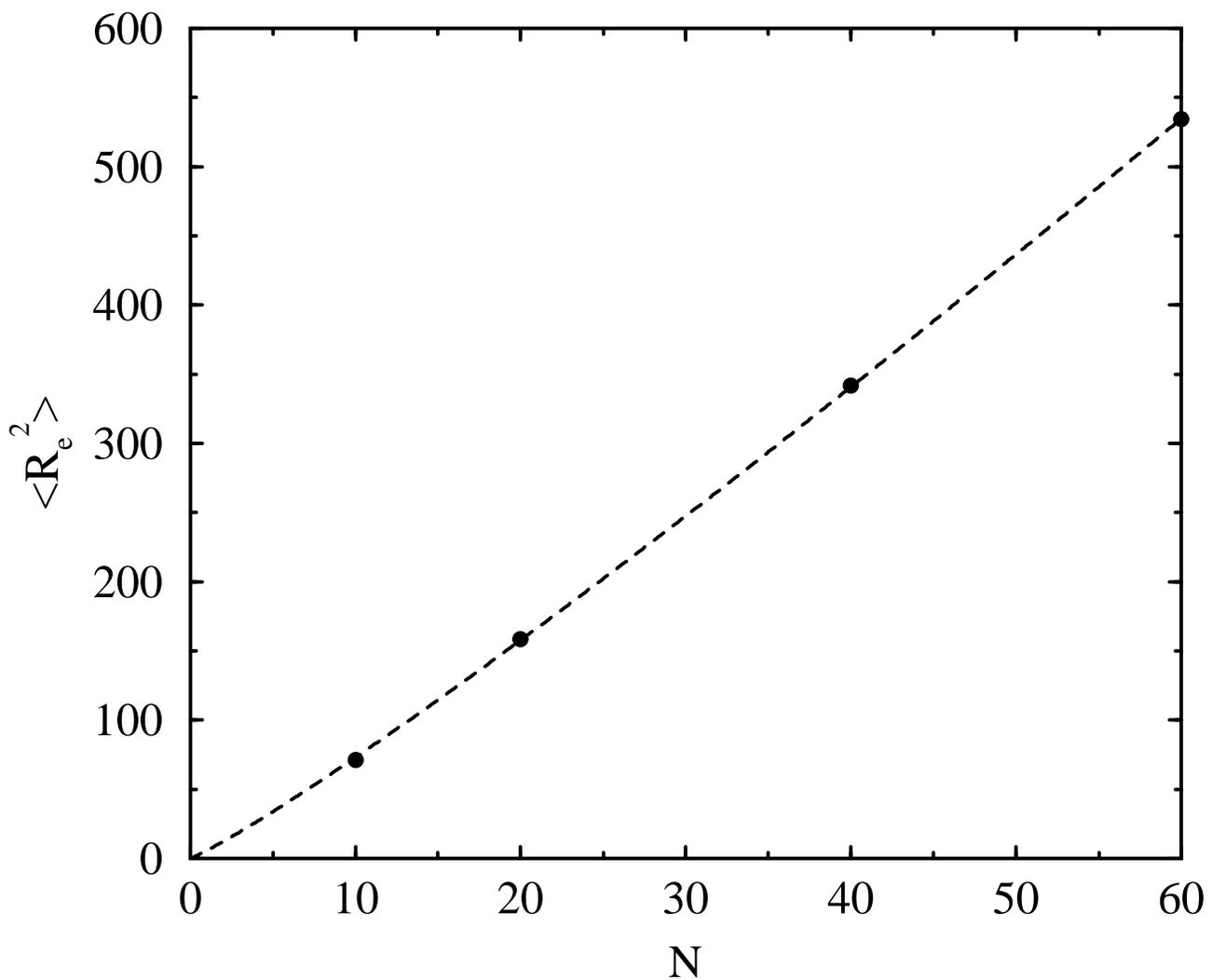}}} 

\caption{The $N$ dependence of the end-to-end distance squared. The
fit is of the form $R_e^2=5.653N^{1.112}$.}

\label{fig:Re2}
\end{figure}
\begin{figure}[h]
\vspace*{0.5 in}
\setlength{\epsfxsize}{19cm}
\centerline{\mbox{\epsffile{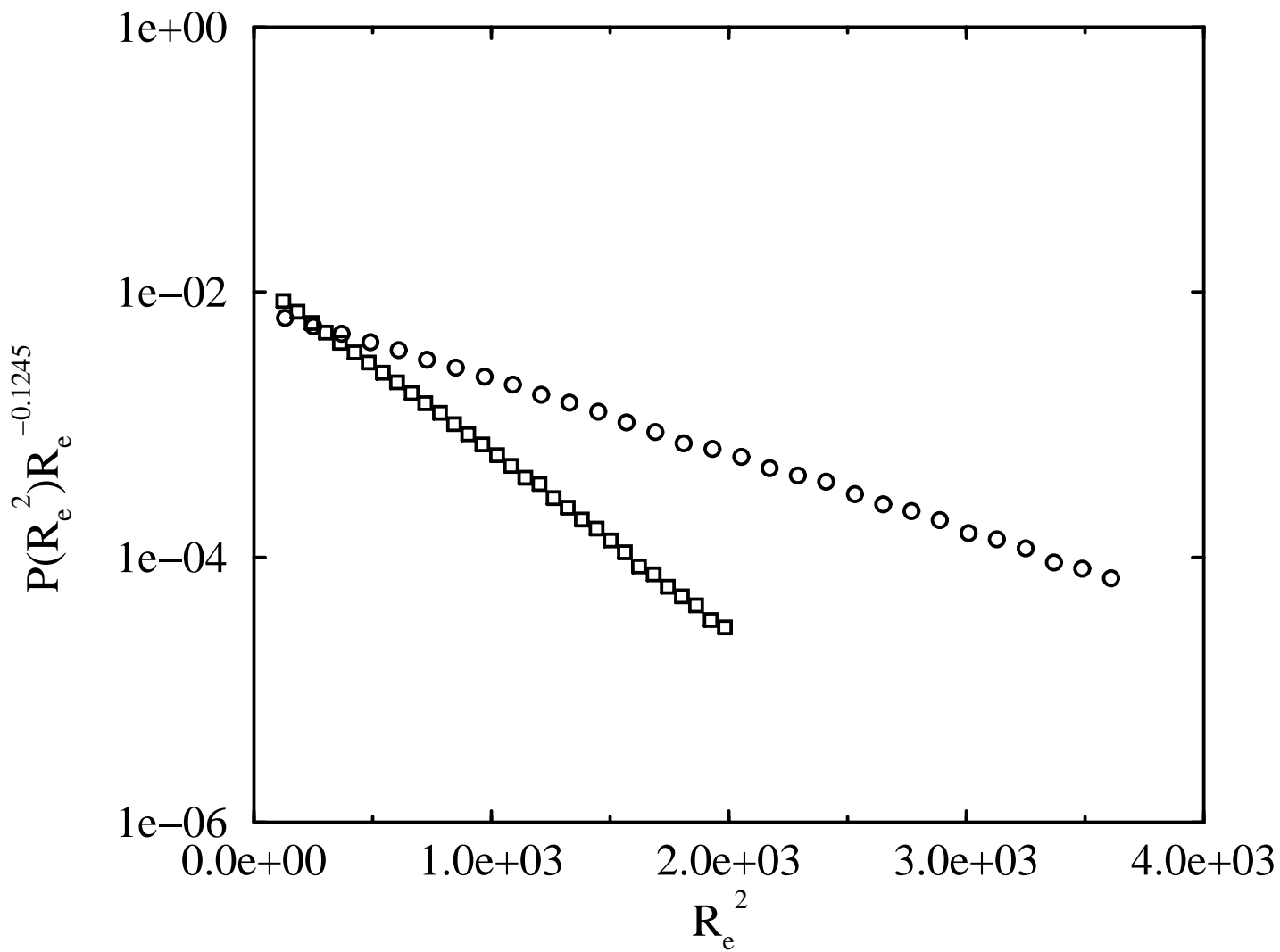}}} 

\caption{The measured function $p(R_e^2)R_e^{-0.1245}$. Fits shows that $\nu>0.5$
implying that the chains are slightly swollen at criticality.}

\label{fig:R2dist}
\end{figure}

Finally we have considered the $N$ dependence of the average squared
end-to-end distance, $R_{ee}^2(N)$ at criticality. This quantity has
been conjectured to scale as \cite{CHERAYIL}:

\begin{equation}
R_{ee}^2(N)\propto N^{2\nu^\prime}
\label{eq:span}
\end{equation}
It was observed that the scaling of $\phi_c$ with $N$ could be
explained if $\nu^\prime=0.46$, and we would like to check this
conjecture. Our results are plotted in figure~\ref{fig:Re2}. Despite
the very limited number of data points, we have attempted to fit this
data to the form Eq.~\ref{eq:span}, with the result
$R_{ee}^2(N)\propto N^{1.11(4)}$. This finding that $\nu^\prime>0.5$
is also supported by a study of the distribution function of
end-to-end distances $p(r)$ at the estimated critical point. A scaling
form for this function (valid for large $N$) may be written
\cite{DESCX}

\begin{equation}
p(r)\propto r^\kappa\exp (-Dr^\delta)
\end{equation}
where $\delta\equiv(1-\nu)^{-1}$, $D$ is a constant and
$\kappa=0.249\pm0.011$ in three dimensions \cite{DESCX}. In
figure~\ref{fig:R2dist} we plot the function $p(r^2)r^{-0.1245}$
against $r^2$ on a logarithmic scale, for the chain length $N=40$ and
$N=60$.  Fits to the data yields estimates $\nu^\prime=0.51$ and
$\nu=0.518$ respectively. Of course it is hard to believe that the
chains are swollen at $T_c(N)$ which is below $\Theta$, given the fact
that for $\phi\rightarrow 0$ at $T=\Theta$, the chains are not
swollen. It is thus possible that the slight deviation from
$\nu^\prime=1/2$ is simply due to corrections to scaling.

\section{Discussion}
\label{sec:discuss}

In summary we have performed a study of the liquid-vapor critical
point of a polymer model for chain lengths up to $N=60$
monomers. Owing to the low acceptance rate for chain transfers it was
not possible to study either very long chains, or very large
systems. Nevertheless we believe that the FSS-based technique we
employed, of matching the measured scaling operator distribution
functions to their fixed point universal forms is considerably more
reliable than the practice of simply extrapolating a power law fit to
coexistence curve data obtained well away from criticality, as has
been the norm in previous simulation studies
\cite{SMIT,SHENG,MACKIE,ESCOBEDO}.  Indeed, while we reproduce the
previous results with regard to the finding that $T_c(N)$ is well
described by a Flory formula, our measured value for the exponent
$x_2=0.37(2)$, is in much closer accord with experiment ($x_2=0.38$)
than previous simulation measurements. It is also interesting to note
that for a similar range of monomeric units $N$, our coarse grained
model seems to be much better at describing the asymptotic limit than
chemically realistic models such as that employed in a the recent
study of Alkanes \cite{SMIT}, which did not even yield a monotonically
decreasing $\phi_c(N)$.

With regard to the critical $N$ dependence of the chain span, our results
suggest (albeit on the basis of a very limited number of data points) that
the exponent $\nu^\prime >0.5$, which, if correct, would imply that the
chains are slightly swollen at criticality---at variance with the suggestion
of Cherayil \cite{CHERAYIL,BISWAS} and L'huiller \cite{LHUILLIER} that
$\nu^\prime\approx 0.46$, which is based on the assumption that the phase
separation occurs when the chains just barely begin to overlap.  We consider
it possible, however, that our estimates for $\nu^\prime$ should be
considered as effective exponents, which exceed the classical value
$\nu^\prime=1/2$ only because of corrections to scaling.  But in any case
there is no evidence that the chains are somewhat collapsed at criticality. 

Finally we remark that there is evidently a need to study longer chain
and larger system sizes in order both to validate the results thus far
obtained and to confirm that the limiting scaling behaviour is being
observed. In view of the low acceptance rates for chain transfers at
large $N$, algorithmic improvements are clearly necessary before this
can be achieved. In this regard, recent improved biased growth
techniques for chain insertion at large $N$ and $\phi$ promise to be
extremely helpful \cite{ALEXAND}. In future work we hope to report on
their application to the present problem.

\subsection*{Acknowledgements}

The authors thank M. Muthukumar for a helpful discussion. NBW
acknowledges financial support from the Max Planck Institut f\"{u}r
Polymerforschung, Mainz. Part of the simulations described here were
performed at the IWR, Universit\"{a}t Heidelberg. Partial support from
the Deutsche Forschungsgemeinschaft (DFG) under grant number Bi314/3-2
and from the Bundesministerium f\"{u}r Bildung, Wissenshaft, Forschung
und Technologie (BMBF) under grant number O3N8008C, is also gratefully
acknowledged.

\end{document}